# *brainlife.io*: A decentralized and open source cloud platform to support neuroscience research

Soichi Hayashi, Bradley A. Caron, Anibal Sólon Heinsfeld, Sophia Vinci-Booher, Brent McPherson, Daniel N. Bullock, Giulia Bertò, Guiomar Niso, Sandra Hanekamp, Daniel Levitas, Kimberly Ray, Anne MacKenzie, Lindsey Kitchell, Josiah K. Leong, Filipi Nascimento-Silva, Serge Koudoro, Hanna Willis, Jasleen K. Jolly, Derek Pisner, Taylor R. Zuidema, Jan W. Kurzawski, Kyriaki Mikellidou, Aurore Bussalb, Christopher Rorden, Conner Victory, Dheeraj Bhatia, Dogu Baran Aydogan, Fang-Cheng F. Yeh, Franco Delogu, Javier Guaje, Jelle Veraart, Steffen Bollman, Ashley Stewart, Jeremy Fischer, Joshua Faskowitz, Maximilien Chaumon, Ricardo Fabrega, David Hunt, Shawn McKee, Shawn T. Brown, Stephanie Heyman, Vittorio Iacovella, Amanda F. Mejia, Daniele Marinazzo, R. Cameron Craddock, Emanuale Olivetti, Jamie L. Hanson, Paolo Avesani, Eleftherios Garyfallidis, Dan Stanzione, James Carson, Robert Henschel, David Y. Hancock, Craig A. Stewart, David Schnyer, Damian O. Eke, Russell A. Poldrack, Nathalie George, Holly Bridge, Ilaria Sani, Winrich A. Freiwald, Aina Puce, Nicholas L. Port, and Franco Pestilli.

**Competing interests.** The authors declare no competing financial interests.

**Corresponding authors.** Franco Pestilli pestilli@utexas.edu

**Contribution**. S.H. implemented the brainlife.io services. B.C. wrote the data analysis code, performed large-scale experiments, and prepared the figures and associated text. A.S.H. improved and implemented some of the services. F.J., C.C., C.C.A., D.H., D.S., D.P., L.K., J.L., C.R., F.N.S., H.W., J.J., Z.T., K.J., S.K., N.A., V.C., B.D., A.D.B., F.D. G.J., S.H., provided assets. All authors edited the manuscript. F.P. invented, designed, and directed *brainlife.io*, wrote the paper, and designed all the experiments, and figures. *Shared first author's contribution.

## ABSTRACT

Neuroscience research has expanded dramatically over the past 30 years by advancing standardization and tool development to support rigor and transparency. Consequently, the complexity of the data pipeline has also increased, hindering access to FAIR data analysis to portions of the worldwide research community. *brainlife.io* was developed to reduce these burdens and democratize modern neuroscience research across institutions and career levels. Using community software and hardware infrastructure, the platform provides open-source data standardization, management, visualization, and processing and simplifies the data pipeline. brainlife.io automatically tracks the provenance history of thousands of data objects, supporting simplicity, efficiency, and transparency in neuroscience research. Here brainlife.io's technology and data services are described and evaluated for validity, reliability, reproducibility, replicability, and scientific utility. Using data from 4 modalities and 3,200 participants, we demonstrate that brainlife.io's services produce outputs that adhere to best practices in modern neuroscience research.

## INTRODUCTION

Over the last 30 years, neuroimaging research has dramatically expanded our ability to study the structure and function of the living human brain, leading to major advancements in understanding brain-related health and disease [1–4]. Today, neuroimaging modalities and techniques span multiple data types (e.g., magnetic resonance imaging [MRI], positron emission tomography [PET], functional near-infrared spectroscopy [fNIRS], electro-encephalography [EEG], and magnetoencephalography [MEG]), and have increased the feasibility of large-scale, population-level, data collection efforts.[1,5,6] At the same time, the field of neuroimaging has attracted a large and ever-growing community of researchers [7,8]. Furthermore, a process of adopting FAIR principles of data stewardship (Findability, Accessibility, Interoperability, and Reusability[9]), data standardization, open science methods, and increased data size, has been gaining grounds and in turns increasing requirements for rigorous and transparent data analysis and reporting. However, such approaches require significant additional technological support, posing new challenges to many researchers. We refer to these challenges as the burdens of neuroscience (**Fig. 1**).

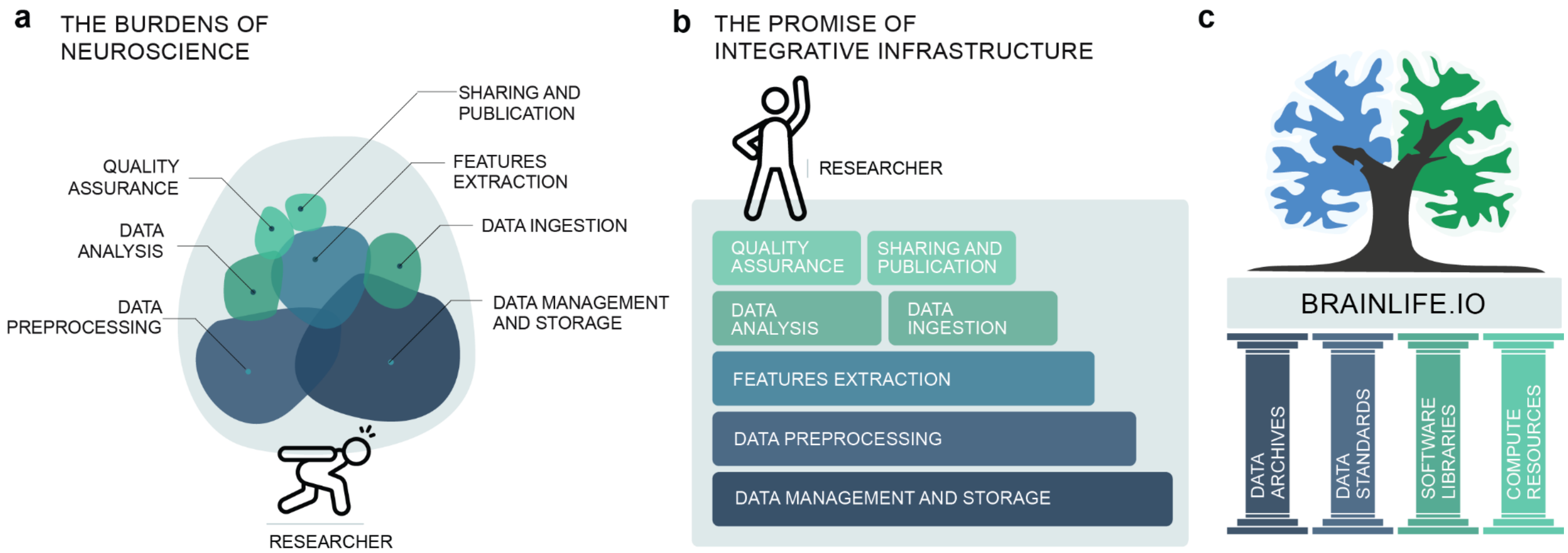

**Figure 1. The burdens of neuroscience. a.** A figurative representation of the current major burdens of performing neuroimaging investigations. **b.** Our proposal for integrative infrastructure that coordinates services required to perform FAIR, reproducible, rigorous, and transparent neuroimaging research thereby lifting the burden from the researcher. **c.** *brainlife.io* rests upon the foundational pillars of the open science community such as data archives, standards, software libraries and compute resources. Panels **a** and **b** adapted from *Eke et al.* (2021).

Datasets are growing in size, in large part because they support scientific rigor and reproducibility. Research on the reproducibility of scientific findings indicates that limited sample sizes might have hindered the validity of early, foundational results in hypothesis-driven cognitive neuroscience research,[10–16] but reproducibility issues can be found in biological science,[17,18] psychology,[12]data science, and computational methods,[19,20] cancer biology,[21], and artificial intelligence.[13,22,23]This is largely because small sample sizes increase the probability of reporting spurious effects as statistically significant.[1,24] Recent findings also make the case for increasing sample sizes into the thousands when research focuses on discovery science.[5] Notable examples of large-scale data sharing within neuroscience and neuroimaging include the Human Connectome Project (HCP),[25] the Cambridge Centre for Ageing and Neuroscience study (Cam-CAN),[26,27] the Adolescent Brain Cognitive Development (ABCD) study,[28,29] the UK-Biobank,[30] the Healthy Brain Network (HBN),[31] the Pediatric Imaging Neurocognition and Genetics (PING) study,[32] the Natural Scene Dataset [33] and the thousands of individual brain datasets deposited on OpenNeuro.org.[34] These data-sharing projects not only serve the needs of the neuroscience community with demonstrated impact [35], but also the incoming generation of AI research.[36–38] However, larger datasets generally entail greater complexity as well. The use of datasets so unprecedented in size requires a substantial scaling up of resources and technical skills, and this in turn results in significant barriers to entry.

Traditionally, neuroimaging researchers have collected a few hours of neuroimaging data on a few dozen subjects and analyzed it using laboratory computers and a single tool-kit or programming environment, often created in-house. Current studies, by contrast, may require the analysis of hundreds (if not thousands) of hours of data, with an accompanying move of data away from individual laboratory computers toward high-performance computing clusters and cloud systems requiring multiple steps and a variety of scripting and programming languages (e.g., Unix/Linux shell, Python, MatLab, R, C++). The complexity of neuroimaging data pipelines and code development stacks have increased concomitantly.[39,40] To help ensure the reproducibility and rigor of scientific results, the neuroimaging community has also developed data standards[41] and software libraries for data processing and analysis (FSL, Freesurfer, Nibabel, MRTrix, DIPY, DSI-STudio).[42–68] More recently prebuilt data processing pipelines that combine software from multiple libraries into unified partially preconfigured steps have been also developed [69–73]. These pipelines advance data processing standardization but still leave many choices of parameters to users and often require technical input data formats.

As a result of all this progress for data and tools, neuroimaging researchers carry the burden of having to piece together and track multiple processes, such as data ingestion and standardization, storage, and management, preprocessing and feature extraction, all while also attending to tracking quality control, analyses, and publication (**Fig. 1a**). Publication of results requires compliance with the FAIR principles which, though well explained in theory, are often challenging to implement in practice. Submission of manuscripts often necessitates new analyses at a later date, by which point software and data versions may have changed, and data might have been removed from compute clusters or local servers. Existing approaches for managing these steps require manual tracking of data and code versions, along with advanced technical skills.[40,74] Currently, there exists no efficient technology to help piece together and keep track of all of these (ever-changing) technology and data requirements.

As the resources necessary to participate fully in modern neuroscience research have grown, barriers to entry and funding have risen as well. Smaller universities, teaching colleges, undergraduate students, and other settings that lack the resources to support significant investments in infrastructure and training are at a meaningful disadvantage. Lack of resources and infrastructure is a key gap identified in surveys pertaining to both the adoption of FAIR neuroscience [75] and the conduct of neuroscience research in low- and medium-income countries [76,77]. Without added support, FAIR neuroscience might evolve with an ever-increasing bias towards high-resourced teams, institutions, and countries. Such an outcome would not only decrease representation and diversity but would slow scientific progress. In support of simplicity, efficiency, transparency, and equity in big data neuroscience research, our team has developed a community resource, *brainlife.io* (**Fig. 1b**). The *brainlife.io* platform stands on the foundational pillars of the neuroimaging community and the mission of open science (**Fig. 1c**). *brainlife.io* provides free and secure reproducible neuroscience data analysis. *brainlife.io*'s technology works for researchers serving automated tracking of data provenance, preprocessing steps, parameter sets, and analysis versions. Our vision for *brainlife.io* is that of a trusted, interoperable, and integrative platform connecting global communities of software developers, hardware providers, and domain scientists via cloud services.

In the remainder of this article, we describe the technology and utilization of *brainlife.io*. After that, we present the results of our evaluations of the effectiveness of the technology. Experiments focused on the four axes of scientific transparency: external validity, reliability, reproducibility, and replicability. Finally, we demonstrate the platform's potential for scientific utility in identifying human disease biomarkers.

## RESULTS

### Platform architecture

*brainlife.io* is a ready-to-use and ready-to-expand platform. As a ready-to-use system, it allows researchers to upload and analyze data from MRI, MEG, and EEG systems. Data are managed using a secure warehousing system that follows an advanced governance and access-control model. Data can be preprocessed and visualized using version-controlled applications (hereafter referred to as Apps) compliant with major data standards (the Brain Imaging Data Structure, BIDS[41]). As a ready-to-expand system, software developers may contribute or modify existing Apps guided by standard methods and documentation describing how to write Apps (github.com/brainlife/abcd-spec and brainlife.io/docs). The platform uses a combination of opportunistic computing and publicly funded resources [78–80] that are functionally integrated and can be available for use by a particular project or team of researchers. Computing resource managers can also register computer servers and clusters on *brainlife.io* to make them available either to individual users or projects or to the larger community of *brainlife.io* users (**Fig. 2a** and **Fig. S2a**). The **Supplemental Platform architecture** provides an extended description of the technology. The platform is available to any type of researcher from students to faculty researchers, either without cost (through opportunistic use of freely contributed resources) or with performance guarantees (through the use of dedicated hardware or payment for use of cloud resources).

Brainlife.io was founded via an initial investment from the U.S. BRAIN Initiative via a National Science Foundation, followed by support from the National Institutes of Health, the Department of Defense, the Kavli Foundation, and the Wellcome Trust. The platform's geographically distributed computing and storage systems are securely hosted by national supercomputing centers and funded by a combination of institutional, national, and international awards (see **Fig. S2**). As of this paper, the Texas Advanced Computing Center, Indiana University Pervasive Technology Institute, Pittsburgh Supercomputing Center, San Diego Supercomputing Center, and the University of Michigan Advanced Research Computing Technology Services have supported the project. The distributed platform is connected with and depends on other major infrastructure and software projects such as OpenNeuro.org, osiris.org, DataLad.org, BIDS, Freesurfer, FSL, nibabel, dipy.org, repronim.org, DSI-Studio, jetstream-cloud.org, frontera-portal.tacc.utexas.edu, access-ci.org, and INCF.org.

The architecture of *brainlife.io* is based on an innovative, microservices-based approach, including authentication, preprocessing, warehousing, event handling, and auditing. This architecture allows automated and decentralized data management and processing. Microservices are handled by the meta-orchestration workflow system Amaretti (**Fig. 2a**,**b**, and **Table S1**). Amaretti can deploy computational jobs on high-performance compute clusters and cloud systems. This allows the utilization of publicly-funded supercomputers and clouds [80], as well as commercial clouds, such as Google Cloud, AWS, or Microsoft Azure.

Data management on *brainlife.io* is centered around Projects and supported by a databasing and warehousing system (github.com/brainlife/warehouse). Projects are the "one-stop-shop" for data management, processing, analysis, visualization, and publication (**Fig. S3c**). Projects are created by users and are private by default, but can also be made publicly visible inside the *brainlife.io* platform. A project can be populated with data using several options (**Fig. 2d**). Several major archives and data repositories are currently docked by *brainlife.io*[74] (see **Fig. 2b**). Noticeable examples are OpenNeuro.org[34] and the Nathan-Kline data-sharing project.[81–83] Datasets can be imported seamlessly into *brainlife.io* Projects by using either the portal brainlife.io/datasets [74] (see **Video S2** and **Video S3**), the standardization tool brainlife.io/ezbids (see **Table S1** and **Video S6**) or a dedicated Command Line Interface (CLI).

Data processing on *brainlife.io* utilizes an object-oriented and micro workflows service model. Data objects are stored using predefined formats, Datatypes, that allow automated App concatenation and pipelining (**Fig. 2c**; brainlife.io/Datatypes). Apps and Datatypes are the key components of a system that work together to allow automated processing and provenance tracking for millions of data objects. Apps are composable processing units written in a variety of languages using containerization technology.[84,85] Apps are smart, and can automatically identify, accept, or reject datasets before processing (**Fig. 2** and **Fig. S2b**). Community-developed data visualizers are served by *brainlife.io* to support quality control (see **Table S1**). Six new data visualizers have been developed and released as part of the project (**Table S1** and **Video S7**). Whenever possible, Datatypes are made compatible with BIDS.[41] BIDS Apps can be easily made into *brainlife.io* Apps and multiple examples exist already brainlife.io/apps.

The data workflow on *brainlife.io* simplifies the complexity of the modern neuroimaging processing pipeline into two steps, akin to Google's MapReduce algorithm.[86] An initial *map step* preprocesses data objects asynchronously and in parallel using Apps, so as to extract features of interest (such as functional activations, white matter maps, brain networks, or time series data; **Fig. 2d**). During the *map step*, Datatypes and Apps are synchronized and moved to available compute resources automatically. Apps process data objects in parallel across study participants in a Project. The *map step* is followed by a *reduce step,* wherein features extracted using Apps are made available to pre-configured Jupyter notebooks[87,88] served on the platform to perform statistical analysis, machine-learning applications, and generate figures. Indeed, all statistical analyses and figures in this paper are available in accessible *Jupyter Notebooks* (see **Table S2**). *brainlife.io's* data workflow makes it possible to integrate large volumes of diverse neuroimaging Datatypes into simpler sets of brain features organized into *Tidy data* structures [89] (**Fig. S3c**).

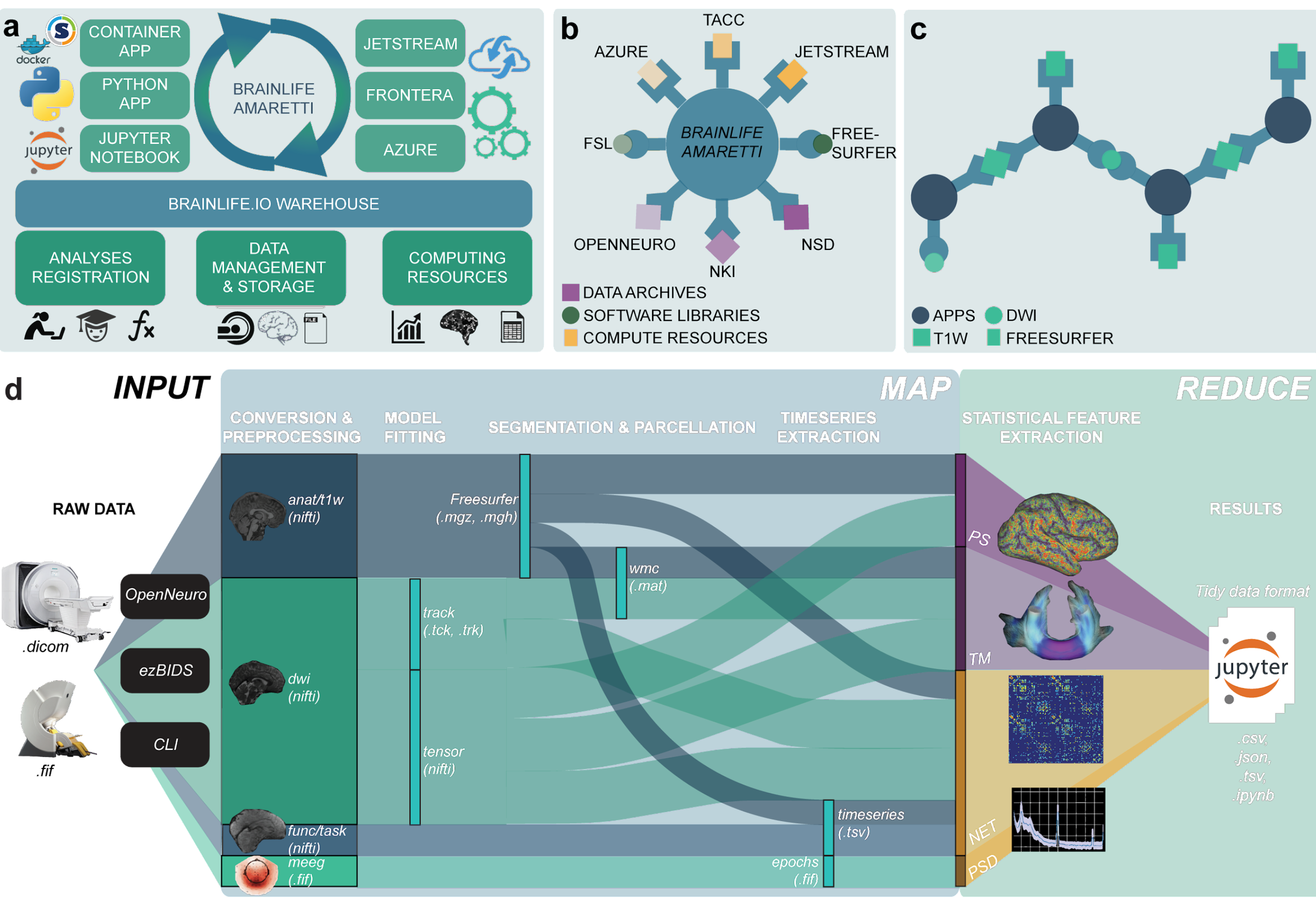

**Figure 2. The *brainlife.io* platform concepts, architecture, and approach. a.** *brainlife.io*'s Amaretti links data archives, software libraries, and computing resources. Specifically, 'Apps' (containerized services defined on GitHub.com) are automatically matched with data stored in the 'Warehouse' with computing resources. Statistical analyses can be implemented using Jupyter Notebooks. **b.** *brainlife.io* provides efficient docking between data archives, processing apps, and compute resources via a centralized service. **c.** Apps use standardized Datatypes and allow "smart docking" only with compatible data objects. App outputs can be docked by other Apps for further processing. **d.** *brainlife.io's Map step* takes MRI, MEG and EEG data and processes them to extract statistical features of interest. *brainlife.io's reduce step* takes the extracted features and serves them to Jupyter Notebooks for statistical analysis. PS: parc-stats Datatype; TM: tractmeasures Datatype; NET: network Datatype; PSD: power-spectrum density Datatype. CLI: Common Line Interface.

A key technological innovation developed for *brainlife.io* is the ability to automatically track all actions performed by platform users on Datatypes and Apps. The platform captures data object IDs, Apps versions, and parameter sets so as to track the full sequence of steps from data import to analysis and publication. A graph describing provenance metadata for each Datatype can be visualized using the provenance visualizer or downloaded (see Fig. S3d and Video S10). A shell script is automatically generated to allow the reproduction of full processing sequences (Video S11). Finally, a single record containing data objects, Apps, and *Jupyter Notebooks* used in a study can be made publicly available outside the platform bundled into a single record addressed by a unique Digital Objects Identifier (DOI) [90]. Whereas all other existing systems provide users with technology to track analysis steps manually or require the use of coding, *brainlife.io* tracks automatically and do not require coding nor user actions to generate a record of everything done by a user for data analysis. This automation technology lowers the barriers of entry and democratizes FAIR, reproducible large-scale neuroimaging data analysis.

## Platform evaluation

In the following section, we evaluate the utility of *brainlife.io*. To do so, we first present the level of engagement with the platform by the growing community of users. After that, we describe the results of experiments on the robustness and validity of the platform. A detailed description of each section below describing each App and step used can be found in the corresponding **Supplemental Platform evaluation** section.

### Platform utilization

*brainlife.io* is developed following the FAIR principles. It is available worldwide and supports thousands of researchers. First made accessible in Spring 2018, its utilization and assets have grown steadily (**Fig. 3** and **Fig. S2c** and **S4**). At the time of this writing, over 2,341 users across 43 countries have created a *brainlife.io* account. Over 1,542 of these have been active users (**Fig. 3a**). Over 3,439 data management Projects have been created, and a community of developers has implemented over 530 data processing Apps. Over 270 TBs of data have been stored and processed using *brainlife.io,* for a total of 1,097,603 hours of compute time.

Researchers ranging from undergraduate students to faculty use *brainlife.io* (**Fig. 3b**), and analyses span the full range of the neuroimaging data lifecycle. The most frequently used Apps pertained to diffusion tractography (22%), model fitting (15%), and anatomical ROI generation (12%). Community-developed software libraries provided the foundations for data processing, including Nibabel, Freesurfer, FSL, DIPY, MRTrix, the Connectome Workbench, and MNE-Python. Terabytes of data have been uploaded (72%) or imported from OpenNeuro.org (22%), the Nathan-Kline Institute data sharing projects (3%[31,81,83]), and other sources. This degree of world-wide platform access highlights the global need for technology like *brainlife.io* (see **Fig. S2e**). More details can be found in **Supplemental platform utilization**.

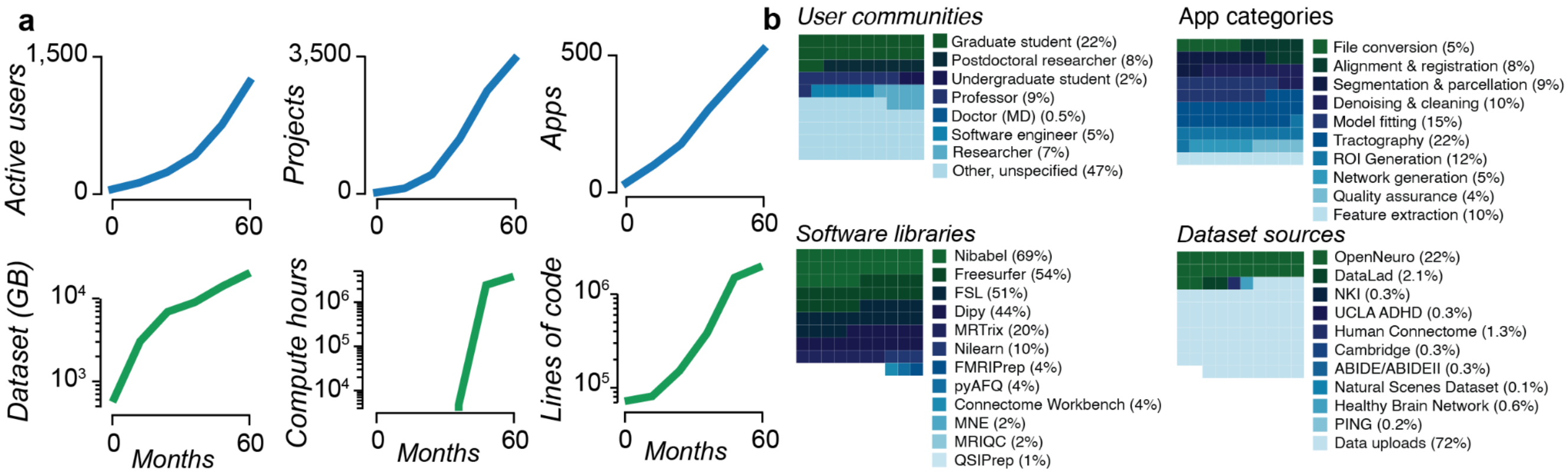

**Figure 3. *brainlife.io* impact (2018-2022). a.** *Top left.* Number of users submitting more than 10 jobs per month. *Top middle.* Number of projects over time. *Top right.* Number of Apps over time. *Bottom left.* Data storage across all Projects. *Bottom middle.* Compute hours across all Projects (data only available 6 months post project start). *Bottom right*. Lines of code in the top 50 most-used Apps. **b.** *Top left.* User communities. *Top right.* App categories. *Bottom left.* Percent of total jobs launched with the software library installed (percentage for jobs of top 50 most-used Apps). *Bottom right.* Datasets sources. See also Fig. S2c for a world-wide distribution of the researchers that have accessed *brainlife.io.*

### Platform testing

Experiments were performed to demonstrate the ability of the platform to provide accurate data processing and analysis at scale. The experiments focused on the four axes of scientific transparency: data processing external validity (DPEV), reliability, reproducibility, and replicability.[91,92] Four data modalities (sMRI, fMRI, dMRI, MEG) were evaluated using, among others, the test-retest $HCP_{TR,}$ [93] the Cam-CAN,[27] the HBN,[31] and the ABCD[28] datasets. In total, data from over 3,200 participants across 12 datasets were processed. Extracted brain features included cortical parcel volumes, white matter tract profilometry, functional and structural network properties, functional gradients, and peak alpha frequency (**Fig. 4**). Over 193,000 data objects and 22 Terabytes of data were generated for the experiments. A detailed description of the experiments below can be found in the **Supplemental platform testing** section. The *brainlife.io* Apps used for the experiments are reported in **Table S3**. Post-processing analyses were performed using *brainlife.io*-hosted Jupyter Notebooks (see **Table S2**).

Data processing external validity (DPEV) was defined as the ability of data processed on *brainlife.io* to accurately reflect brain properties proficiently processed by other teams. DPEV was estimated for four data modalities (sMRI, dMRI, fMRI, and MEG) and five brain features (brain areas volumes, major white matter tracts fractional anisotropy, resting state functional connectivity, resting-state function gradients, and MEG peak alpha frequency). Features values obtained using *brainlife.io* Apps were compared against data preprocessed by data originators, specifically the HCP consortium or Cam-CAN project team (**Fig. 4**, Fig. S4**d,e,h**). Cortical area volume estimates on 148 parcels were obtained using *brainlife.io* Apps and compared to corresponding estimates provided by the HCP consortium (**Fig. 4a**; $r_{validity}$=0.98, $rmse_{validity}$=570.54mm$^3$). Fractional anisotropy (FA) in 61 white matter tracts was estimated using the raw and minimally preprocessed HCP$_{TR}$ dMRI data (**Fig. 4b**; $r_{validity}$=0.95, $rmse_{validity}$=0.018). Functional connectivity estimates between 117$^2$ nodes-pairs [94] were compared between raw and minimally preprocessed HCP$_{TR}$ dMRI data (**Fig. 4c**; $r_{validity}$=0.89, $rmse_{validity}$=0.12). In addition, functional gradients [95,96] were computed on 400 nodes estimated on raw and minimally processed HCP$_{TR}$ fMRI data (**Fig. 4d**; $r_{validity}$=0.59, $rmse_{validity}$=0.036). Finally, the peak alpha frequency values were compared between Cam-CAN and *brainlife.io* processed MEG data (**Fig. 4e**; $r_{validity}$=0.94, $rmse_{validity}$=0.30 Hz). Overall, the results show strong similarity in feature estimates between data processed on *brainlife.io* versus those processed by external groups (functional gradients demonstrated the lowest validity and data processing-type dependency based on fMRI preprocessing procedures [97]).

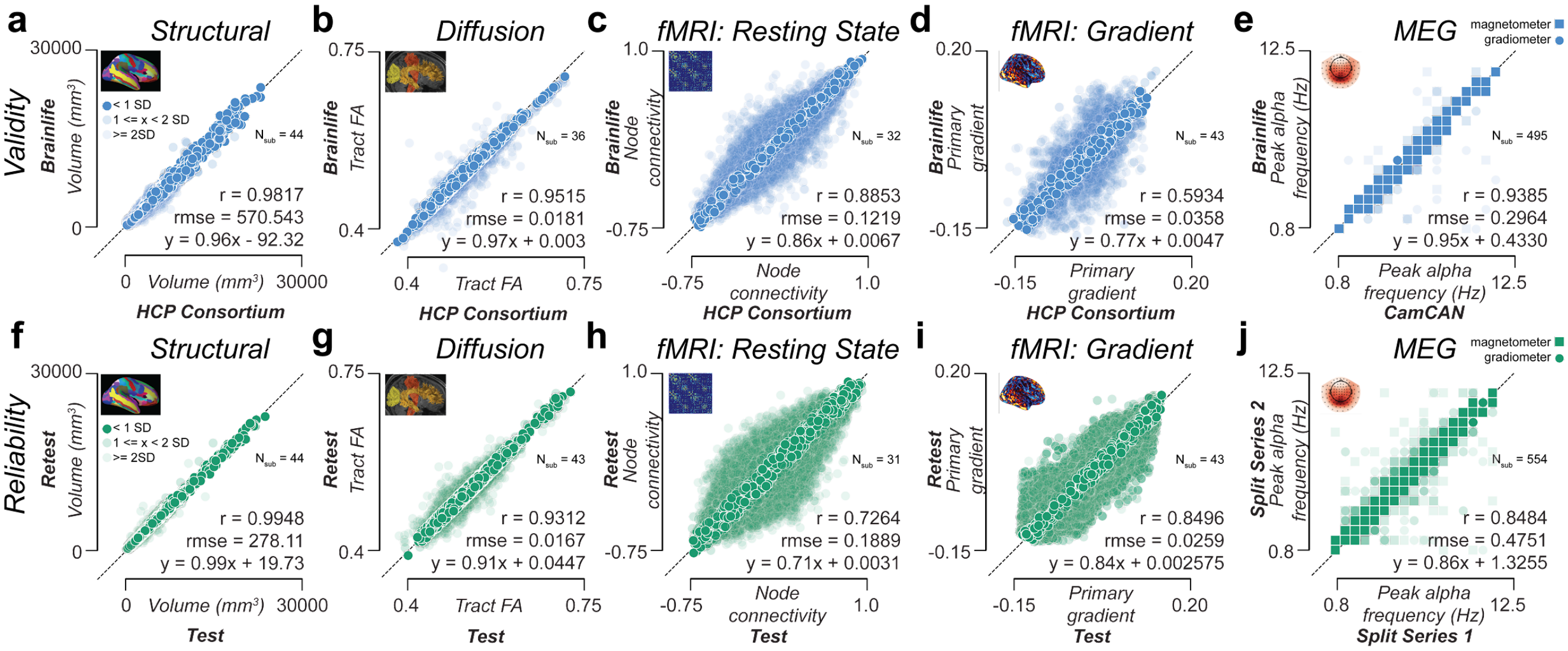

**Figure 4. Data processing validity and reliability analysis*. Top row:*** Validity measures derived using the HCP Test-Retest data. Each dot corresponds to the ratio for a given subject between data preprocessed and provided by the HCP Consortium vs data preprocessed on *brainlife.io* in a given measure for a given structure. Pearson's correlation (*r*), root mean squared error (*rmse*), and a linear fit between the test and retest results were calculated. **a.** Parcel volume (mm$^3$). **b.** Tract-average fractional anisotropy (FA). **c*.** Node-wise functional connectivity (FC). **d*.** Primary gradient value derived from resting-state fMRI. **e.** Peak frequency (Hz) in the alpha band derived from MEG. Data from magnetometer sensors are represented as squares, and data from gradiometer sensors are represented as circles. ***Bottom row*: ** Test-retest reliability measures derived from derivatives of the HCP$_{TR}$ dataset generated using *brainlife.io*. Each dot corresponds to the ratio between a test-retest subject and a given measure for a given structure. Pearson's correlation (*r*), root mean squared error (*rmse*), and a linear fit between the test and retest results were calculated. **f.** Parcel volume (mm$^3$). **g.** Tract-average fractional anisotropy (FA). **h*.** Node-wise functional connectivity (FC). **i*.** Primary gradient value derived from resting-state fMRI. **j.** Peak frequency (Hz) in the alpha band derived from MEG using the Cambridge (Cam-CAN) dataset. Data from magnetometer sensors are represented as squares, and data from gradiometer sensors are represented as circles. Dark colors represent data within +/-1 standard deviation (SD. 50% opacity represents data within 1-2 SD. 25% opacity represents data outside 2 SD. *A representative 5% of data presented in **c**, **d**, **h**, **i**.

Data processing reliability (DPR) was defined as the ability to produce highly similar results on *test* and *retest measurements* within a study participant. DPR was estimated for the four data modalities and five brain features used above to estimate DPEV. Brain features estimated using *brainlife.io* Apps on *test* and *retest measurements* (HCP$_{TR}$ dataset) or median splits data (Cam-CAN MEG) were compared. Reliability estimates of brain area volumes, major tracts FA, networks FC, functional gradients, and Peak Alpha Frequency were obtained (see **Fig. 4f-i** and associated supplemental text). DPR varied between $r_{reliability}$=0.99 and 0.73, with sMRI and dMRI demonstrating the highest reliability ($r_{reliability}$=0.99, 0.93, respectively). See also Fig. S4f-g,i for estimates on additional brain features and **Table S4** for a full report of all correlation values obtained in all brain features. The results show strong reliability of most of all the pipelines with the fMRI reliability being lowest, this is consistent with previous reports [98]. We also performed computational reproducibility (CR) experiments (see **Fig. S4j-n** and associated text). These experiments demonstrated the similarity in estimates produced by *brainlife.io* Apps when used twice to process the same dataset. Given the use of containerization technology for the Apps, this test was expected to return high correlation values. Indeed, all correlations were above 0.99, demonstrating high consistency. These experiments demonstrate the ability of the platform to conduct valid, reliable, and reproducible data processing and analysis at scale across multiple data modalities and brain features.

### Platform utility for scientific applications

Next, we evaluated the platform's potential to support scientific findings. To do so, we evaluated whether data processed using *brainlife.io*'s Apps contained meaningful patterns. We used over 1,800 participants from three datasets: PING (Pediatric Imaging, Neurocognition, Genetics), HCP$_{s1200}$, (HCP Young Adult 1,200), and Cam-CAN. Data were collected across ages, but age ranges differed in each dataset (i.e., 3-20 years for PING, 20-37 years for HCP$_{s1200}$, and 18-88 years for Cam-CAN). The lifelong trajectory was plotted for multiple brain features (e.g., volumes of brain parts, FA of major tracts, network properties. MEG peak frequency, etc; **Fig. 5**). The collated age range spanned 7 decades. Features were combined using *brainlife.io*'s Jupyter Notebooks.

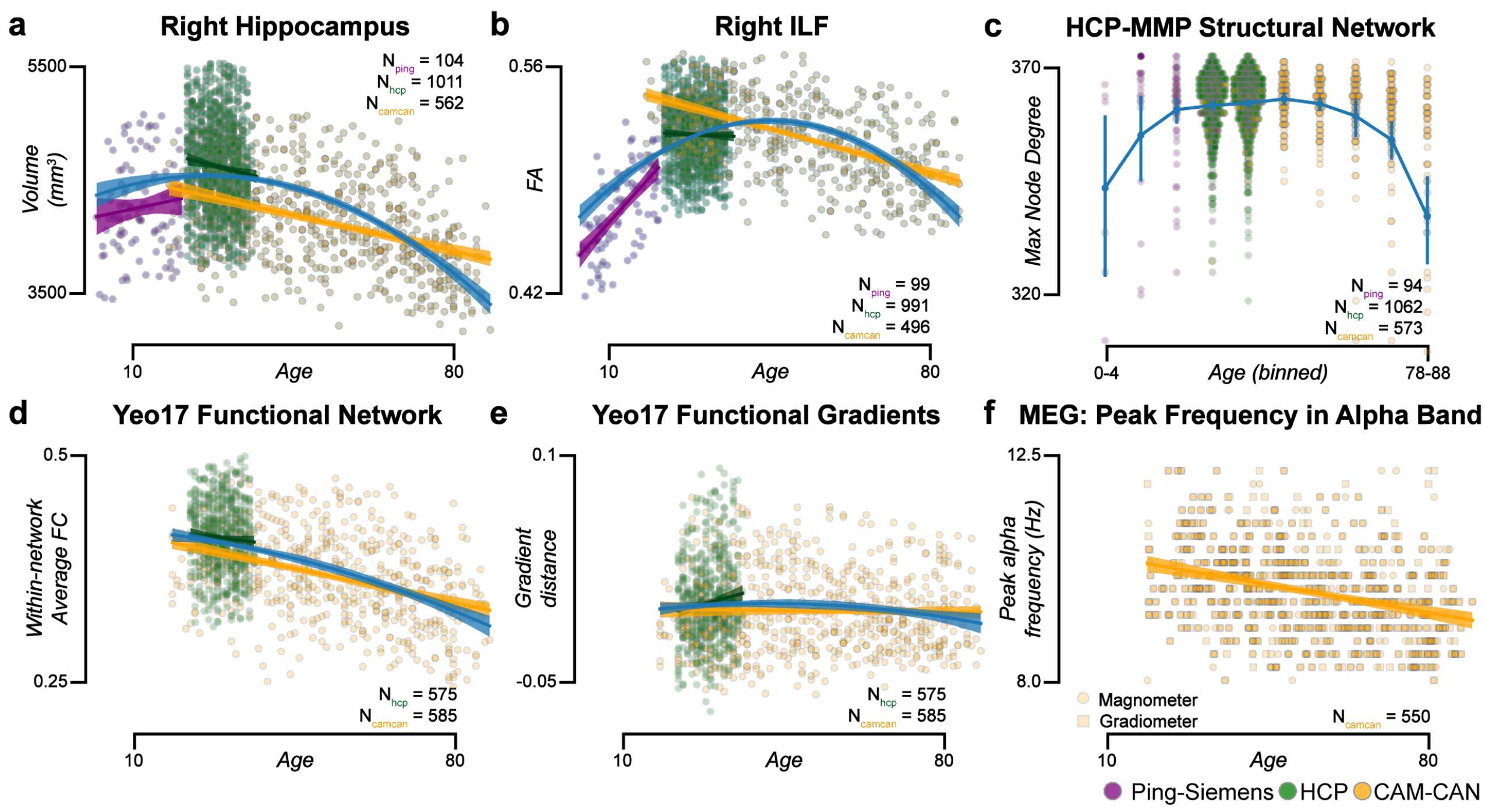

**Figure 5. Lifelong brain maturation estimated across datasets.** Relationship between subject age and **a.** Right hippocampal volume, **b.** Right inferior longitudinal fasciculus (ILF) fractional anisotropy (FA), **c***. maximum node degree of density network derived using the *hcp-mmp* atlas, **d*.** Within-network average functional connectivity (FC) derived using the Yeo17 atlas, **e**. Functional gradient distance for visual resting state network derived from the Yeo17 atlas, and **f.** Peak frequency in the alpha band derived from magnetometer (squares) and gradiometers (circles) from MEG data. These analyses include subjects from the PING (*purple*), HCP$_{1200}$ (*green*), and Cam-CAN (*yellow*) datasets. Linear regressions were fit to each dataset, and a quadratic regression was fit to the entire dataset (blue). * All points in **c**, and **d** are presented. See also **Fig. S5** and **Supplemental platform utility for scientific applications**.

Multiple reports have shown inverted U-shaped lifelong trajectories across data modalities.[99–103] We plotted brain features derived for each data modality (sMRI, dMRI, fMRI, and MEG) as a function of age across datasets (**Fig. 5**). Six exemplary lifelong trajectories are shown (additional features are reported in **Fig. S5**). For each data modality, a quadratic model was fit across all three datasets between 3 and 88 years of age: $y_{feature} = ax_{age}^2 + bx_{age} + c$, ($R^2$=0.152 ± 0.0773 s.d.). Mean quadratic term (*a*) across all data modalities was negative (-0.0514 ± 0.111 s.d.), demonstrating the expected inverted U-shape trajectory. Results show that, by automatically analyzing data using *brainlife.io* Apps, it is possible to collate across datasets with substantial differences in data acquisition parameters and signal-to-noise profiles. Additional details regarding these experiments can be found in **Supplemental platform utility for scientific applications**.

### Replication and generalization of previous results

We then evaluated the ability of *brainlife.io* to replicate previous results and generalize findings across datasets. A more detailed description and additional experiments can be found in **Supplemental replication and generalization**. First, we tested *brainlife.io*'s ability to replicate the results of three previous studies. A negative correlation between cortical thickness and tissue orientation dispersion (ODI; $r_{original}$ =-0.46) has been reported in the $HCP_{s1200}$ dataset.[104] *brainlife.io* Apps were created to estimate cortical thickness and ODI and analyze $HCP_{s1200}$ dataset. A negative relationship between cortical thickness and ODI was estimated, replicating the original study (**Fig. 6a**; $r_{HCP\text{-}brainlife}$ = -0.43 vs. $r_{original}$). More examples of replications can be found in **Fig. S6a,b**.

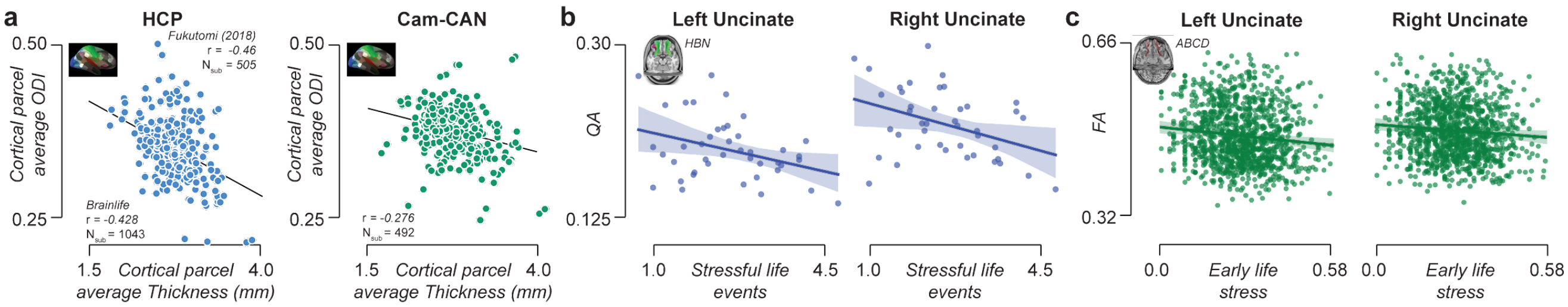

**Figure 6. Replication of previous studies using brainlife.io. a.** Average cortical hcp-mmp parcel thickness ($N_{struc}$ = 322) compared to parcel orientation dispersion index (ODI) from the NODDI model mapped to the cortical surface (*inset*) of the HCP S1200 dataset ($N_{sub}$ = 1,043) and Cam-CAN ($N_{sub}$ = 492) dataset compared to the parcel-average cortical thickness. **b.** Stressful life events obtained from Negative Life Events Schedule (NLES) survey from Healthy Brain Network participants ($N_{sub}$ = 42) compared to Uncinate-average normalized Quantitative Anisotropy (QA). Mean linear regression (*blue line*) fits and standard deviation (*shaded blue*). **c.** Early life stress was obtained from multiple surveys collected from ABCD participants ($N_{sub}$ = 1,107) compared to Uncinate-average Fractional Anisotropy (FA). Linear regression (*green line*) fits the data with standard deviation (*shaded green*).

Second, the generalization of the original findings to a different dataset was tested in three ways. The first test was run using the cortical ODI estimated in the Cam-CAN dataset. A negative trend of about half the magnitude of the original was estimated (**Fig. 6a**; $r_{Cam\text{-}CAN\text{-}brainlife}$ = -0.28 vs. $r_{original}$). The result generalizes the original results and the reduced effect in a new dataset is consistent with reports on the reproducibility of scientific findings.[12] The second generalization test focused on the reported relationship between life stressors and white matter structural organization of the uncinate fasciculus (UF; r=-0.057).[105] Two datasets were used to extend the finding to new data, i.e., HBN and ABCD. The number of negative life events (Negative Life Events Schedule; NLES) in the HBN dataset was correlated with subjects' quantitative anisotropy (QA) in the right- and left-hemisphere UF. Results show a negative correlation similar in magnitude as found in the original study (**Fig. 6b** $r_{HBN_LEFT}$ = -0.35, p-value < 0.05; $r_{HBN_RIGHT}$ = -0.39, p-value < 0.05). The third and final attempt at the generalization of the same result was made using the ABCD dataset. Early life stress was estimated as a composite score of traumatic life events, environmental and neighborhood safety, and the family conflict subscale of the Family Environment Scale.[29] A negative relationship between UF FA and the composite score was estimated in the left- and right-UF (**Fig. 6c** $r_{ABCD_LEFT}$ = -0.12, p-value < 0.001; $r_{ABCD_RIGHT}$ = -0.09, p < 0.01). Overall, these results demonstrate both the robustness of the original results and the potential of *brainlife.io* services to detect meaningful associations in large, heterogeneous datasets.

### Example applications to detecting disease

The final two tests evaluated the platform's ability to identify human disease biomarkers. Data from individuals with a sports-related concussion, eye disease (Choroideremia and Stargardt's disease), and matched controls were used (**Fig. 7**). A detailed description of the experiments can be found in **Supplemental to detecting disease**. It has been reported that concussion can alter brain tissue both in cortical and deep white matter tracts.[106] We set out to measure the difference in cortical white matter tissue in concussed and matched controls. FA was estimated from data collected within 24-48 hours post-concussion. The distribution of FA in the superior temporal sulcus (STS) is reported (**Fig. 7a**). One representative athlete showed strong post-concussive symptoms and low STS cortical FA (red). The result demonstrates the potential of *brainlife.io* processed data to report meaningful changes in brain tissue following a concussion.

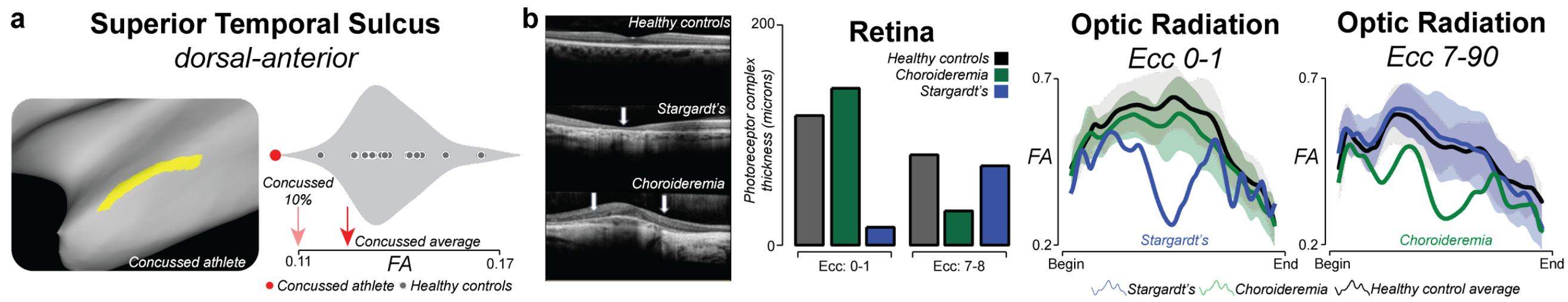

**Figure 7. Using brainlife.io to identify and characterize clinical populations from healthy controls. a.** Fractional anisotropy (FA) values were estimated within the superior temporal sulcus (da: dorsal anterior) from 20 healthy athlete controls (*gray distribution*) and 10 concussed athletes. Average FA, 10% low FA, and the lowest FA value across all concussed athletes were measured (*red arrows and dot*). **b**. Retinal OCT images from healthy controls (*top row*), Stargardt's disease patients (*middle row*), and Choroideremia patients (*bottom* row). From these images, photoreceptor complex thickness was measured for each group (Controls: gray; Choroideremia: green; Stargardt's: blue) in two distinct areas of the retina: the fovea (eccentricities 0-1 degrees) and the periphery (eccentricities 7-8 degrees). In addition, optic radiations carrying information for each area of the retina were segmented and FA profiles were mapped. Average profiles with standard error (shaded regions) were computed. One Stargardt and one Choroideremia participant were each identified as having FA profiles that deviated from both healthy controls and the opposing retinal disorder.

Changes in the white matter of the optic radiation (OR) as a result of eye disease have been reported.[107–111] We set out to test the ability of *brainlife.io* Apps to detect similar changes in the OR white matter tissue in two eye diseases for which OR white matter changes have not previously been reported. Individuals with Stargardt's disease (a deterioration of the retina initiating in the central fovea), and Choroideremia (retinal deterioration initiating in the visual periphery), were compared to healthy controls. Retina photoreceptor complex thickness was estimated in the fovea and peripheral using optical coherence tomography (0-1 and 7-90 degrees of visual eccentricity, respectively; **Fig. 7b**). Choroideremia patients showed photoreceptor complex thickness comparable to healthy controls in the fovea, but deviated in the periphery (**Fig. 7b**). The trend was opposite for Stargardt's patients. *brainlife.io* Apps were developed to automatically separate OR bundles projecting to different visual eccentricity in cortical area V1. Average FA profiles for each patient group and controls were estimated for OR fibers projecting to the fovea or periphery.[112] [113,114] Results show a reduction in FA in the component of the OR projecting to the fovea (but not the periphery) in Stargardt's patients (**Fig. 7b**, blue), and the opposite pattern (OR fibers projecting to the periphery had lower FA than controls) in Choroideremia patients (**Fig. 7b**, blue). These results demonstrate the ability of the platform technology to detect disease biomarkers.

### A new approach to facilitate quality control at scale

*brainlife.io* offers a unique quality assurance (QA) approach to ensure processed data has the quality necessary to serve large user bases. *Reference ranges are* often used in vision science to provide a reference for a measurement, [115] and a similar approach was integrated within the *brainlife.io* data processing interface. To test it, the mean, first, and second SD were estimated (via multiple Apps) for four brain features (tractmeasures, parc-stats, networks, PSD) using the $HCP_{s1200}$, Cam-CAN, and PING datasets. For each of the four brain features, the estimated mean and estimated s.d. (referred to here as *Reference ranges*) are automatically calculated on the

*brainlife.io* platform. That is, when a researcher uses an App to estimate one of the four features, the values of the researcher's dataset are automatically overlaid on top of the mean, first, and second s.d. marks provided as a reference by *brainlife.io.* In this way, the mean and variability can be used by researchers to efficiently judge whether a recently processed dataset returned appropriate values. For example, reference datasets can be used to detect outlier data (**Fig. 8a-d**). Example reference datasets for four Datatypes are in **Fig. 8e** and an example of platform interfaces reporting these reference datasets is shown in **Fig. S8**. A detailed description of the approach used in this section can be found in **Supplemental to quality control at scale**. These reference ranges are an additional source for quality assurance, alongside other options for QA such as online data visualization, the automated generation of images and plots from the processed data as well as the detailed technical reports from major BIDS Apps such as fMRIprep, QSIPrep, MRIQC, Freesurfer [69,70,72,116].

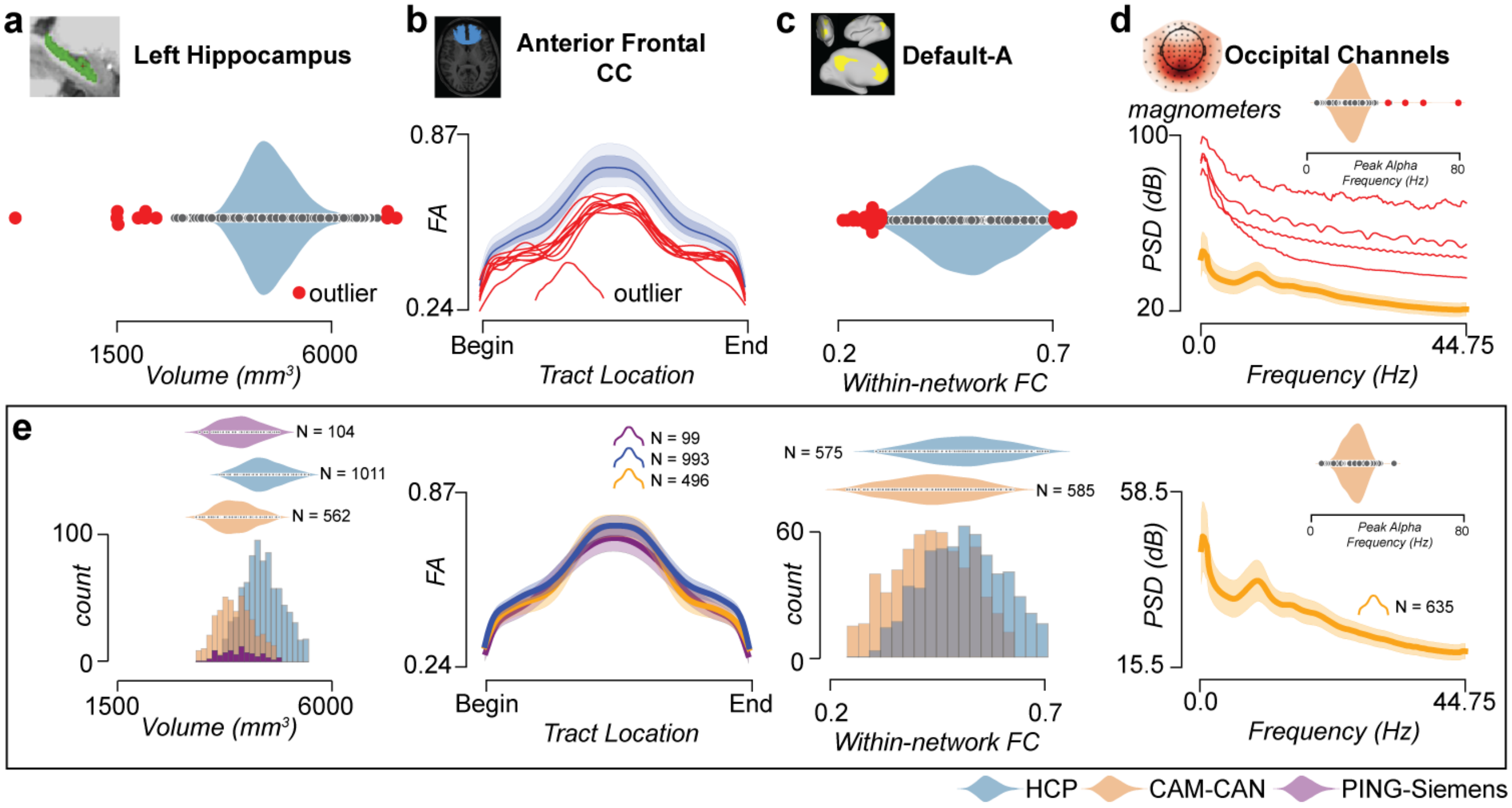

**Figure 8. Reference datasets for quality assurance.** Example workflow for building normative reference ranges for multiple derived statistical products (cortical parcel volume, white matter tract profilometry, within-network functional connectivity, and power-spectrum density (PSD)). **a.** Cortical volumes of the left hippocampus from HCP participants. Red dots indicate outlier data points. **b.** Average fractional anisotropy (FA) profiles (blue line) plotted with two standard deviations (shaded regions). Red lines indicate outlier profiles. **c.** Within-network functional connectivity for the nodes within the Default-A network using the Yeo17 atlas. Red dots indicate outlier data points. **d.** Average PSD from occipital channels using magnetometer sensors from Cam-CAN participants with one standard deviation (shaded regions). Red lines indicate outlier participants. Peak alpha frequency distribution was also computed, and outliers were detected (inset). **e.** Normative reference distributions for each derived statistical product across the PING (*purple*), HCP (*blue*), and Cam-CAN (*orange*) datasets. These distributions have had outliers removed. An example of the brainlife visualization for reference datasets can be found in **Fig. S8**.

## DISCUSSION

The *brainlife.io* platform was developed with public funding to promote the progress of brain science and education and to enable discovery and improve health. The platform connects researchers with publicly available datasets, analysis code, data archives, and compute resources. *brainlife.io* is an end-to-end, turnkey data analysis platform that provides researchers interested in the brain with services for data upload, management, visualization, preprocessing, analysis, and publication–all integrated within a unique cloud environment and web interface. The platform uses opportunistic computing and publicly-funded resources for storage and computing,[78–80] but it can also use popular commercial clouds. The goal is to advance the democratization of big data neuroscience by lowering the barriers of entry to multimodal data analysis, network neuroscience, and large-scale analysis, all opportunities historically limited to a paucity of highly-skilled, high-profile research teams.[39,99,117–122] The platform supports a rigorous and transparent scientific process spanning the research data lifecycle from after data collection to sharing[123] and automatically tracks complex sequences of interactions between researchers, Apps, analysis notebooks, and data objects to support reproducibility. The FAIR data principles for data stewardship and management [9] are generally used as guidelines for any data-centric project. Recently, it has been proposed that a modern definition of neuroscience data should extend beyond measurements and data to include metadata and software for analysis and management. [123] Each research asset on *brainlife.io* (i.e., data derivatives, analysis software, and software services, as handled by the platform) is aligned with the FAIR data principles (see **Supplement on brainlife.io and the FAIR principles**). The following discussion will include descriptions of the resources available for getting started on brainlife.io, applications of *brainlife.io* to educational settings, the platform's strict data governance principles, increasing "data gravity" via *brainlife.io*, potential expansion of the platform, and the platform's current limitations.

The *brainlife.io* project provides multiple resources for App developers, computing resource managers, and neuroscience researchers to learn to use the platform or contribute to the project. A comprehensive overview of the platform and tutorials for getting started with developing Apps or using the platform can be found in the integrated documentation (brainlife.io/docs), as well as on a YouTube Channel that provides tutorials and demonstrations of concepts (youtube.com/@brainlifeio). A public slack channel is used for managing user communications, requests, feedback, and operations (brainlife.slack.com). Users can also ask questions to developers and the community using the topic 'brainlife' on neurostars.org and adding GitHub issues. Finally, a quarterly community engagement and outreach newsletter is sent to all users, and a Twitter account (@brainlifeio) informs the wider community on critical events and connects to information relevant to the project.

*brainlife.io* and its user community are highly engaged in providing innovative training and education opportunities for the next generation of students, postdocs, and clinicians interested in the intersection between neuroscience, data science, and information. The platform allows new students and educators to access many complex data files and analysis methods with minimal overhead. Educators have started using *brainlife.io* to teach neuroscience and data science concepts in the classroom, and courses have been organized in Europe, the USA, Canada, and Africa. These courses introduce basic concepts and teach students how to perform neuroimaging investigations without the requirement of programming or computing expertise. The skills that can be learned using the platform include data preprocessing, quality assurance, and statistical analyses. Integrative data management and analysis provide opportunities for educators and students in under-resourced institutions or countries to perform research and teach neuroscience with hands-on experience.

The project leadership and advisory team recognize the importance of ensuring that data processing workflows are ethically responsible, legally compliant, and socially acceptable. Indeed, data governance is considered an integral part of data processing. Data governance is defined as the principles, procedures, technologies, and policies that ensure acceptable and responsible processing of data at each stage of the data life cycle.[123] It comprises the management of the availability, usability, integrity, quality, and security of data.[123] The data governance policies, processes, and technologies within *brainlife.io* cover three key elements: people, processes, and technologies. A comprehensive set of advanced security measures and protocols guarantee that only authorized individuals have access. These measures include end-to-end encrypted communication, strict access control, and support for multi-factor authentication. Datasets uploaded by users using brainlife.io/ezBIDS are pseudonymized,[124] (i.e. direct identifiers are removed) at upload. The platform interface provides fields for project managers to add Data Use Agreements (DUA) in alignment with the nature and context of their data. The platform even provides template DUAs describing data users' responsibilities and liabilities, including becoming the data controller (the person who controls the purposes and means of processing the data). These governance mechanisms comply with available regulations and mandates, such as the European Union's General Data Protection Regulation (GDPR) and the Health Insurance Portability and Accountability Act (HIPAA) in the United States, which require that personal data be stored and managed in a secure and compliant manner. Cloud systems are designed to provide the level of protection necessary to ensure the privacy and confidentiality of research participants. Finally, the incoming changes to data deposition and sharing mandates (such as that recently released by the National Institutes of Health in the United States[125,126]) are likely to increase the workload for neuroscience researchers. The *brainlife.io* publication records are compatible with the NIH data sharing mandates (for privacy, sharing, and preservation), and the platform is registered on fairsharing.org, datacite.org, datasetsearch.research.google.com, and nitric.org.

Data gravity is the ability of datasets to attract utilization[127] Neuroimaging research within the larger neuroscience field has led the way in increasing data gravity. A long and growing list of tools orchestrated under a general label of open science are being developed to support and facilitate data utilization and access. These tools can be divided into four primary categories: software library, data archives and database systems, data standards, and computing platforms.[40] The data archives and systems closest to *brainlife.io* are the INDI,[128,129] OpenNeuro.org,[34] DANDI,[130] BossDB,[131] DataLad,[74] NITRC,[132] PING,[32] Can-CAM,[27] the Brain/MINDS project,[133] and LORIS.[134] The web services most related to the current work are NeuroQuery,[135] NeuroScout,[136] CBRAIN,[137] NeuroDesk,[138] XNAT,[139] NEMAR,[140] EBRAINS [141], LONI, [142,143], the International Brain Lab data Instratructure [144], COINSTAC [145] and CONP [146]. Most projects are open-source and provide various degrees of data access. *brainlife.io* end-to-end integrated environment that brings researchers from raw data to Jupyter Notebooks and Tidy data tables while tracking data provenance automatically is unique. But many other projects exist and given the fast-growing landscape of neuroinformatics projects, we collected a table listing the major ones (see **Table S5**). The International Neuroinformatics Coordinating Facility also provides a list of major projects incf.org/infrastructure-portfolio. brainlife.io is one of the approved resources, as it complies with the INCF requirement for FAIR infrastructure. The ability of the platform to utilize data from multiple modalities (MEG, EEG, MRI) is a unique feature, connecting neuroimaging research sectors that have been historically siloed. However, we envision additional opportunities for expanding the types of data managed by the platform, fostering further data integration. For example, other data modalities could be mapped to *brainlife.io* Datatypes, and the mechanism for data Integration with metadata capture toolkits [147] and data models [148] would provide additional facilitation for the analysis domains of data currently not covered by the BIDS standard.

Improving the platform's automation and interoperability is part of the vision and sustainability plan. For example, despite the best efforts of App developers, errors occur (see **Fig. S3d**). Currently, researchers only have simple interfaces that report technical output logs and error messages when Apps fail to process data, and parsing these messages requires expertise. Users are required to either contact the *brainlife.io* team or parse the error logs themselves. Planned improvements to *brainlife.io*'s error reporting interfaces will help users understand the sources of errors and find solutions. In addition to error identification, identifying the optimal set of processing steps or parameter sets at the beginning of a project can prove challenging. In addition, currently, researchers identify the optimal data processing steps by looking at existing documentation or videos. In the future, mechanisms that automatically identify processing steps can be implemented to suggest to researchers optimal ways to process their data (e.g. given what other researchers might have already implemented on the platform). Finally, improving connection with major archives and platforms such as OpenNeuro.org, DANDI, NeuroScout, NeuroDesk, and neurosynth.org, would contribute to implementing the vision of a global interoperable ecosystem for a FAIR, accessible, and democratized neuroscience.

In summary, the capabilities of brainlife.io are unique, open, accessible, and expandable. The expansion of instrument capabilities in neuroimaging has in the last 30 years revolutionized our ability to collect data about the brain and brain function. As the landscape of neuroscience big-data projects is only expected to grow in the coming years, moving research data management and computing to cloud platforms will become not just a brilliant option, but a serious requirement. Compliance with mandates for data privacy and sharing will ultimately require researchers to move data management and processing to secure and professionally managed to compute and storage systems. Our goal for *brainlife.io* is to facilitate this process and thereby revolutionize the ability to rigorously and reliably make use of the wealth of data now available to understand brain function, leading to new cures for brain disease. In so doing, *brainlife.io* will also make cutting-edge datasets and analysis resources more accessible to students and researchers from traditionally underrepresented groups in high-, medium- and low-income countries.

## ONLINE METHODS AND MATERIALS

**Data collection approval.** Multiple experiments were performed by individuals at various institutions using the platform. Experiments were approved by the local institutional review boards (IRB), and only the personnel approved for a specific study accessed the data in private projects on brainlife.io. Some of the secondary data usages were deemed IRB-exempt.

**Data sources.** Multiple openly available data sources were used for examining the validity, reliability, and reproducibility of brainlife.io Apps and for examining population distributions. All information regarding the specific image acquisitions, participant demographics, and study-wide preprocessing can be found in the following publications [27,28,31,149–153]. Some data sources are currently unpublished. For these, the appropriate information is provided.

**Validity, reliability, reproducibility, replicability, developmental trends, & reference datasets**
*Human Connectome Project (HCP; Test-Retest, s1200-release)* [149]. Data from these projects were used to assess the validity, reliability, and reproducibility of the platform. They were used to assess the abilities of the platform to identify developmental trends in structural and functional measures, and they were used to generate reference datasets. Structural data (sMRI): The minimally-preprocessed structural T1w and T2w images from the Human Connectome Project (HCP) from 1066 participants from the s1200 and 44 participants from the Test-Retest releases were used. Specifically, the 1.25 mm 'acpc_dc_restored' images generated from the Siemens 3T MRI scanner were used for all analyses involving the HCP. For most examinations, the already-processed Freesurfer output from HCP was used. Diffusion data (dMRI): To assess the validity of preprocessing on brainlife.io, the unprocessed dMRI data from 44 participants from the HCP Test dataset was used. For reliability and all remaining analyses, the minimally-preprocessed diffusion (dMRI) images from 1,066 participants from the s1200 and 44 participants from the Test-Retest releases from the 3T Siemens scanner were used. All processes incorporated the multi-shell acquisition data. Functional data (fMRI): For validation, the unprocessed resting-state functional MRI (fMRI) from 44 participants from the HCP Test dataset was compared to the minimally-preprocessed BOLD data provided by HCP. For reliability and all other analyses, the minimally-preprocessed BOLD data from 1,066 participants from the s1200 and 44 participants from the Test-Retest releases from the 3T Siemens scanner were used.

*The Cambridge Centre for Ageing and Neuroscience (Cam-CAN)* [27]. The data from this project were used to assess the validity, reliability, and reproducibility of the platform and to assess the abilities of the platform to identify developmental trends of structural and functional measures, and to generate reference datasets. Structural data (sMRI): The unprocessed 1mm isotropic structural T1w and T2w images from 652 participants from the Cambridge Centre for Ageing and Neuroscience (Cam-CAN) study were used. Diffusion data (dMRI): The unprocessed 2mm isotropic diffusion (dMRI) images from 652 participants from the Cambridge Centre for Ageing and Neuroscience (Cam-CAN) study were used. Functional data (fMRI): The 3mm x 3mm x 4mm unprocessed resting-state fMRI images from 652 participants from the Cambridge Centre for Ageing and Neuroscience (Cam-CAN) study were used. Electromagnetic data (MEG): The 1000 Hz resting-state filtered and unfiltered datasets from 652 participants from the Cambridge Centre for Ageing and Neuroscience (Cam-CAN) study were used.

**Developmental trends & reference datasets**
*Pediatric Imaging, Neurocognition, and Genetics (PING)* [32]. The data from this project were used to assess the abilities of the platform to identify developmental trends of structural measures and to generate reference datasets. Structural data (sMRI): The unprocessed 1.2 x 1.0 x 1.0 mm structural T1w and the 1.0 mm isotropic T2w images from 110 participants from the *Pediatric Imaging, Neurocognition, and Genetics (PING)* study were used. Diffusion data (dMRI): The unprocessed 2mm isotropic diffusion (dMRI) images from 110 participants from the *Pediatric Imaging, Neurocognition, and Genetics (PING)* study were used.

**Replicability datasets**
*Adolescent Brain Cognitive Development (ABCD)* [28,29]. Structural data (sMRI): The unprocessed 1mm isotropic structural T1w and T2w images from a subset of 1,877 participants from the Adolescent Brain Cognitive Development (ABCD release-2.0.0) study were used. Diffusion data (dMRI): The unprocessed 1.77mm isotropic diffusion (dMRI) images from a subset of 1877 participants from the Adolescent Brain Cognitive Development (ABCD release-2.0.0) study were used. A single diffusion gradient shell was used for these experiments (b=3000s/msec$^2$). Research approved by the University of Arkansas IRB (#2209425822).

*Healthy Brain Network (HBN)* [31]. The data from this project were used to assess the abilities of the platform to replicate previously published findings via the assessment of the relationship between microstructural measures mapped to segmented uncinate fasciculi and self-reported early life stressors. Research approved by the University of Pittsburgh IRB (#PRO17060350). Structural data (sMRI): The 0.8 mm isotropic structural T1w images from 42 participants from the Healthy Brain Network (HBN) study were used. Diffusion data (dMRI): The unprocessed 1.8 mm isotropic diffusion (dMRI) images from 42 participants from the CitiGroup Cornell Brain Imaging Center site of the Healthy Brain Network (HBN) study were used. Research approved by the University of Pittsburgh IRB (#PRO17060350).

*UPENN-PMC [154].* The data from this project were used to assess the abilities of the platform to replicate previously published findings via the assessment of the performance of an automated hippocampal segmentation algorithm. All procedures were conducted under the approval of the Institutional Review Board at the University of Texas at Austin. Structural data (sMRI): The T1w and T2w data were provided within the Automated Segmentation of Hippocampal Subfields (ASHS) atlas[154].

**Clinical-identification datasets**
*Indiana University Acute Concussion Dataset.* The data from this project were used to assess the abilities of the platform to identify clinical populations via the mapping of microstructural measures to the cortical surface. Neuroimaging was performed at the Indiana University Imaging Research Facility, housed within the Department of Psychological and Brain Sciences with a 3-Tesla Siemens Prisma whole-body MRI using a 64-channel head coil. Within this study, 9 concussed athletes and 20 healthy athletes were included. Research approved by Indiana University (IRB: 906000405). Structural data (sMRI): High-resolution T1-weighted structural volumes were acquired using an MPRAGE sequence: TI = 900 ms, TE = 2.7 ms, TR = 1800 ms, flip angle = 9°, with 192 sagittal slices of 1.0 mm thickness, a field of view of 256 x 256 mm, and an isometric voxel size of 1.0 $mm^3$. The total acquisition time was 4 minutes and 34 seconds. High-resolution T2-weighted structural volumes were also acquired: TE = 564 ms, TR = 3200 ms, flip angle = 120°, with 192 sagittal slices, a field of view of 240 x 256 mm, and an isometric voxel size of 1.0$mm^3$. Total acquisition time was 4 minutes 30 seconds. Diffusion data (dMRI): Diffusion data were collected using single-shot spin-echo simultaneous multi-slice (SMS) EPI (transverse orientation, TE = 92.00 ms, TR = 3,820 ms, flip angle = 78 degrees, isotropic 1.5 $mm^3$ resolution; FOV = LR 228 mm x 228 mm x 144 mm; acquisition matrix MxP = 138 x 138. SMS acceleration factor = 4). This sequence was collected twice, one in the AP fold-over direction and the other in the PA fold-over direction, with the same diffusion gradient strengths and the number of diffusion directions: 30 diffusion directions at b = 1000 s/$mm^2$, 60 diffusion directions at b = 1,750 s/$mm^2$, 90 diffusion directions at b = 2,500 s/$mm^2$, and 19 b = 0 s/$mm^2$ volumes. The total acquisition time for both sets of dMRI sequences was 25 minutes and 58 seconds.

*Oxford University Choroideremia & Stargardt's Disease Dataset.* The data from this project was used to assess the abilities of the platform to identify clinical populations via mapping retinal-layer thickness via OCT and mapping of microstructural measures along optic radiation bundles segmented using visual field information (eccentricity). Neuroimaging was performed at the Wellcome Centre for Integrative Neuroimaging, Oxford with the Siemens 3T scanner. Research approved by the UK Health Regulatory Authority reference 17/LO/1540. Structural data (sMRI)**:** High-resolution T1-weighted anatomical volumes were acquired using an MPRAGE sequence: TI = 904 ms, TE = 3.97 ms, TR = 1900 ms, flip angle = 8°, with 192 sagittal slices of 1.0 mm thickness, a field of view of 174 mm x 192 mm x 192 mm, and an isometric voxel size of 1.0 $mm^3$. The total acquisition time was 5 minutes and 31 seconds. Diffusion data (dMRI): Diffusion data were collected using EPI (transverse orientation, TE = 92.00ms, TR = 3600 ms, flip angle = 78 degrees, 2.019 x 2.019 x 2.0 $mm^3$ resolution; FOV = 210 mm x 220 mm x 158 mm; acquisition matrix MxP = 210 x 210, SMS acceleration factor = 3). This sequence was collected twice, one in the AP fold-over direction and the other in the PA fold-over direction. The PA fold-over scan contained 6 diffusion directions, 3 at b = 0 s/$mm^2$ and 3 at b = 2000 s/$mm^2$, and was used primarily for susceptibility-weighted corrections. The AP fold-over scan contained 105 diffusion directions, 5 at b = 0 mm/$s^2$, 51 at b = 1000 mm/$s^2$, and 49 at b = 2000 mm/$s^2$. The total acquisition time for both sets of dMRI sequences was 7 minutes and 8 seconds.

***General processing pipelines***
**Structural processing.** For the ABCD, Cam-CAN, Oxford University Choroideremia & Stargardt's Disease Dataset, and the Indiana University Acute Concussion datasets, the structural T1w and T2w (sMRI) images (if available) were preprocessed, including bias correction and alignment to the anterior commissure-posterior commissure (ACPC) plane, using A273 and A350 respectively. For PING data, no bias correction was performed but alignment to the ACPC plane was performed using A99 and A116 for T1w and T2w data respectively. For HCP data, this data was already provided. The structural $T_1$-weighted images for each participant and dataset were then segmented into different tissue types using functionality provided by *MRTrix3* (Tournier et al, 2019) implemented as A239. For a subset of datasets, this was performed within the diffusion tractography generation step using A319. The gray- and white-matter interface mask was subsequently used as a seed mask for white matter tractography. The processed structural T1w and T2w images were then used for segmentation and surface generation using the *recon-all* function from Freesurfer[72] (A0). Following Freesurfer, representations of the cortical 'midthickness' surface were computed by spatially averaging the coordinates of the pial and white matter surfaces generated by Freesurfer using the wb_command -surface-cortex-layer function provided by Workbench command for the $HCP_{TR}$, $HCP_{s1200}$, ABCD, Cam-CAN, PING, and Indiana University Acute Concussion datasets. These surfaces were used for cortical tissue mapping analyses. Following Freesurfer and midthickness-surface generation, the 180 multimodal cortical nodes (*hcp-mmp*) atlas and the Yeo 17 (*yeo17*) atlas were mapped to the Freesurfer segmentation of each participant implemented as brainlife.io App A23. These parcellations were used for subsequent cortical, subcortical, and network analyses. In addition, measures for cortical thickness, surface area, volume, and summaries of diffusion models of microstructure were estimated using A383 and A389. To estimate population receptive fields (pRF) and visual field eccentricity properties in the cortical surface in the Oxford University Choroideremia & Stargardt's Disease Dataset, the automated mapping algorithm developed by [155,156] was implemented using A187. To segment thalamic nuclei for optic radiation tracking, the automated thalamic nuclei segmentation algorithm provided by Freesurfer [72] was implemented as A222. Finally, visual regions of interest binned by eccentricity were then generated using AFNI [157] functions implemented in A414. To assess the replicability capabilities of the platform, an automated hippocampal nuclei segmentation app (A262) was used to segment hippocampal subfields from participants within the UPENN-PMC dataset provided within the ASHS atlas.

**Diffusion (dMRI) processing.** *Preprocessing & model fitting:* For a majority of the analyses involving the HCP dataset, the minimally-preprocessed dMRI images were used and thus no further preprocessing was performed. However, to assess the validity of the preprocessing pipeline, the unprocessed dMRI data from the HCP Test dataset, dMRI images were preprocessed following the protocol outlined in [158] using A68. The same app was also used for preprocessing the dMRI images for the ABCD, Cam-CAN, PING, Oxford University Choroideremia & Stargardt's Disease Dataset, the Indiana University Acute Concussion, and HBN datasets. Specifically, dMRI images were denoised and cleaned from Gibbs ringing using functionality provided by *MRTrix3* before being corrected for susceptibility, motion, and eddy distortions and artifacts using FSL's *topup* and *eddy* functions [44,159]. Eddy-current and motion correction was applied via the *eddy_cuda8.0* with the replacement of outlier slices (*i.e. repol*) command provided by FSL [160–163]. Following these corrections, MRTrix3's *dwigradcheck* functionality was used to check and correct for potential misaligned gradient vectors following top-up and eddy [164]. Next, dMRI images were debiased using ANT's *n4* functionality [165] and the background noise was cleaned using MrTrix3.0's *dwidenoise* functionality [166]. Finally, the preprocessed dMRI images were registered to the structural (T1w) image using FSL's *epi_reg* functionality [167–169]. Following preprocessing, brain masks for dMRI data using *bet* from FSL were implemented as A163.

*DTI, NODDI, and q-sampling model fitting.* Following preprocessing, the diffusion tensor (DTI) model [170] and the neurite orientation dispersion and density imaging (NODDI) [171,172] models were subsequently fit to the preprocessed dMRI images for each participant using either A319 or A292 for DTI model fitting and A365 for NODDI fitting. Note, the NODDI model was only fit on the HCP, Cam-CAN, Oxford University Choroideremia & Stargardt's Disease Dataset, and the Indiana University Acute Concussion datasets. For those datasets, the NODDI model was fit using an intrinsic free diffusivity parameter ($d_{\parallel}$) of 1.7x10-3 mm$^2$/s for white matter tract and network analyses, and a $d_{\parallel}$ of 1.1x10-3mm$^2$/s for cortical tissue mapping analyses, using AMICO's implementation[172] as A365. The constrained spherical deconvolution (CSD) (Tournier et al, 2007) model was then fit to the preprocessed dMRI data for each run across 4 spherical harmonic orders (i.e. $L_{max}$) parameters (2,4,6,8) using functionality provided by *MRTrix3* implemented as brainlife.io App A238. For the PING datasets, the CSD model was fit using the same exact code found in A238, but performed using the tractography App A319. For the HBN dataset, the isotropic spin distribution function was obtained by reconstructing the diffusion MRI data with the Generalized q-sampling imaging method [173] using functionality provided by DSI-Studio[66] (A423). Quantitative anisotropy (QA) was then estimated from the isotropic spin distribution function.

*Tractography.* Following model fitting, the fiber orientation distribution functions (fODFs) for $L_{max}$=6 and $L_{max}$=8 were subsequently used to guide anatomically-constrained probabilistic tractography (ACT; Smith et al, 2012) using functions provided by *MRTrix3* implemented as brainlife.io App A297 or A319. For the $HCP_{TR}$, $HCP_{s1200}$, and Oxford University Choroideremia & Stargardt's Disease datasets, $L_{max}$=8 was used. For ABCD and Cam-CAN datasets, $L_{max}$=6 was used. For the HCP, ABCD, Cam-CAN, datasets, a total of 3 million streamlines were generated. For all datasets, a step-size of 0.2 mm was implemented. For the $HCP_{TR}$, $HCP_{s1200}$, ABCD, and Cam-CAN datasets, minimum and maximum lengths of streamlines were set at 25 and 250mm respectively, and a maximum angle of curvature of 35° was used. For the PING dataset, minimum and maximum lengths of streamlines were set at 20 and 220mm respectively, and a maximum angle of curvature of 35° was used.

*White Matter Segmentation and cleaning.* Following tractography, 61 major white matter tracts were segmented for each run using a customized version of the white matter query language (Bullock et al, 2019) implemented as brainlife.io App A188. Outlier streamlines were subsequently removed using functionality provided by Vistasoft and implemented as brainlife.io App A195. Following cleaning, tract profiles with 200 nodes were generated for all DTI and NODDI measures across the 61 tracts for each participant and test-retest condition using functionality provided by Vistasoft and implemented as A361. Macrostructural statistics, including average tract length, tract volume, and streamline count was computed using functionality provided by Vistasoft implemented as A189. Microstructural and macrostructural statistics were then compiled into a single data frame using A397.

*Segmentation of the optic radiation (OR).* To generate optic radiations segmented by estimates of visual field eccentricity in the Oxford University Choroideremia & Stargardt's Disease Dataset, ConTrack [111] tracking was implemented as A252. 500,000 sample streamlines were generated using a step size of 1mm. Samples were then pruned using inclusion and exclusion waypoint ROIs following methodologies outlined in [108,109].

*Segmentation of uncinate fasciculus (UF).* To assess the relationship between Uncinate tract-average quantitative anisotropy (QA) and fractional anisotropy (FA) and Early Life Stressors within two independent datasets (Healthy Brain Network, ABCD), the tract-average QA for the Left and Right Uncinates were computed from 42 participants from the HBN and the tract-average FA were computed from 1107 participants from the ABCD dataset. For the HBN dataset, a full tractography segmentation pipeline was used to preprocess the dMRI data and segment the uncinate fasciculus using A423. Automatic fiber tracking was then performed to segment the uncinate fasciculus using default parameters and templates from a population tractography atlas from the Human Connectome Project [174]. A threshold of 16 mm as the maximum allowed threshold for the shortest streamline distance was then applied to remove spurious streamlines. The whole tract average QA was then estimated. To probe stress exposure within the HBN dataset, we used the Negative Life Events Schedule (NLES), a 22-item questionnaire where participants were asked about the occurrence of different stressful life events. For the questions pertaining to early life stressors, the ABCD dataset was used. The tract-average FA for the Left and Right Uncinates were estimated using procedures described previously, then compared to the participant's life stressors behavioral measures by fitting a linear regression to the data.

*Structural networks:* Following tract segmentation, structural networks were generated using the multi-modal 180 cortical node atlas and the tractograms for each participant using MRTrix3's *tck2connectome*[175] functionality implemented as A395. Connectomes were generated by computing the number of streamlines intersecting each ROI pairing in the 180 cortical node parcellation. Multiple adjacency matrices were generated, including count, density (i.e. count divided by the node volume of the ROI pairs), length, length density (i.e. length divided by the volume of the ROI pairs), and average and average density AD, FA, MD, RD, NDI, ODI, and ISOVF. Density matrices were generated using the *-invnodevol* option[176]. For non-count measures (length, AD, FA, MD, RD, NDI, ODI, ISOVF), the average measure across all streamlines connecting and ROI pair was computed using MRTrix3's *tck2scale* functionality using the *-precise* option[177] and the *-scale_file* option in *tck2connectome*. These matrices can be thought of as the "average measure" adjacency matrices. These files were outputted as the 'raw' Datatype, and were converted to *conmat* Datatype using A393. Connectivity matrices were then converted into the 'network' Datatype using functionality from python functionality implemented as A335.

*Cortical & subcortical diffusion & morphometry mapping.* For the PING, $HCP_{TR}$, $HCP_{s1200}$, Cam-CAN, and Indiana University Acute Concussion datasets, DTI and NODDI (if available) measures were mapped to each participant's cortical white matter parcels following methods found in Fukutomi and colleagues using functions provided by Connectome Workbench[93] implemented as *brainlife.io* App A379. A Gaussian smoothing kernel (FWHM = ~4mm, σ = 5/3mm) was applied along the axis normal to the midthickness surface, and DTI and NODDI measures were mapped using the wb_command -volume-to-surface-mapping function. Freesurfer was used to map the average DTI and NODDI measures within each parcel using functionality from Connectome Workbench using A389 and A483. Measures of volume, surface area, and cortical thickness for each cortical parcel were computed using Freesurfer and A464. Freesurfer was also used to generate parcel average DTI and NODDI measures for the subcortical segmentation (*aseg*) from Freesurfer using A383. Measures of volume for each subcortical parcel were computed using Freesurfer and A272.

**Resting-state Functional (rs-fMRI) preprocessing and functional connectivity matrix generation.** For the $HCP_{TR}$ and Cam-CAN datasets, unprocessed rs-fMRI datasets were preprocessed using fMRIPrep implemented as A160. Briefly, fMRIPrep does the following preprocessing steps. First, individual images are aligned to a reference image for motion estimation and correction using *mcflirt* from FSL. Next, slice timing correction is performed in which all slices are realigned in time to the middle of each TR using *3dTShift* from AFNI. Spatial distortions are then corrected using field map estimations. Finally, the fMRI data is aligned to the structural T1w image for each participant. Default parameters provided by fMRIPrep were used. For a subset of analyses involving the HCP Test and Retest datasets, the preprocessed rs-fMRI datasets provided by the HCP consortium were used. Following preprocessing via fMRIPrep for the volume data, connectivity matrices were generated using the Yeo17 parcellation and A369 and A532. Within-network functional connectivity for the 17 canonical resting state networks was computed by computing the average functional connectivity values within all of the nodes belonging to a single network. These estimates were used for subsequent analyses.

**Resting-state Functional (rs-fMRI) gradient processing.** For the $HCP_{TR}$ and Cam-CAN datasets, unprocessed rs-fMRI data from HCP Test and the Cam-CAN datasets were preprocessed using fMRIPrep implemented as A267. Within this app, the same preprocessing steps are undertaken as in A160, except for an additional volume-to-surface mapping using *mri_vol2surf* from Freesurfer. The surface-based outputs were then used to compute gradients following methodologies outlined in [96] for each participant in the $HCP_{s1200}$, $HCP_{TR}$, and Cam-CAN datasets using A574 using diffusion embedding [178] and functions provided by BrainSpace [179]. More specifically, connectivity matrices were computed from surface vertex values within each node of the Schaffer 1,000 parcellation [180]. Cosine similarity was then computed to create an affinity matrix to capture inter-area similarity. Dimensionality reduction is then used to identify the primary gradients. A normalized-angle kernel was used to create the affinity matrix, from which two primary components were identified. Gradients were then aligned across all participants using a Procrustes alignment and joined embedding procedure [96]. Values from the primary gradient and the cosine distance used to generate the affinity matrices were used for subsequent analyses.

**Magnetoencephalography (MEG) processing.** For some analyses, raw resting-state magnetoencephalography (rs-MEG) time series data from the Cam-CAN dataset was filtered using a Maxwell filter implemented as A476 and median split using A529. For the remainder of the analyses, filtered data provided by the Cam-CAN dataset was used. For all MEG data, power-spectrum density profiles (PSD) were estimated using functionality provided by MNE-Python [181] implemented as A530. Following PSD estimation, peak alpha frequency was estimated using A531. Finally, PSD profiles were averaged across all nodes within each of the canonical lobes (frontal, parietal, occipital, temporal) using [A599](). Measures of power-spectrum density and peak alpha frequency were used for all subsequent analyses.

## DATA AVAILABILITY.

All data derived and described in this paper are made available via the *brainlife.io* platform as "Publications". User data agreements are required for some projects, like data from the HCP, Cam-CAN, PING, ABCD, and HBN datasets. The *Indiana University Acute Concussion Dataset* and the *Oxford University Choroideremia & Stargardt's Disease Dataset* are parts of ongoing research projects and are not being released at this current time. All other datasets are made freely available via the *brainlife.io* platform. See supplementary Table 6 for the brainlife.io/pubs [we have added one example data record (https://doi.org/10.25663/brainlife.pub.40) for the review process <the DOIs for the remaining data records will be added at publication>].

## CODE AVAILABILITY.

As part of the article we are describing a total of 9 platform components. All components are made publicly available open source under MIT License. All the software for the platform components is listed in Supplementary Table 1. In addition, we share the code used for the statistical analyses as Jupyter Notebooks (Supplementary Table 2). Finally, the Apps used and tested in this article are listed in Supplementary Table 3.

## ACKNOWLEDGMENTS.

The *brainlife.io* project, development and operations were supported by awards to Franco Pestilli; U.S. National Science Foundation (NSF) awards 1916518, 1912270, 1636893, and 1734853; U.S. National Institutes of Health awards (NIH) R01MH126699, R01EB030896, and R01EB029272; The Wellcome Trust award 226486/Z/22/Z; A Microsoft Investigator Fellowship; A gift from the Kavli Foundation. Additional funding was provided to support data collection used by the team, research that used *brainlife.io* or infrastructure that supported the platform: NSF awards 2004877 (Sophia Vinci-Booher), 1541335 and 2232628 (Shawn McKee), 1445604 and 2005506 (David Hancock), 1341698 (Michael Norman), 1928224 (Michael Norman), 1445606 (Shawn Brown), 1928147 (Sergiu Sanalevici). NIH awards 1U54MH091657 (HCP data, PIs David Van Essen and Kamil Ugurbil), U01DA041048, U01DA050989, U01DA051016, U01DA041022, U01DA051018, U01DA051037, U01DA050987, U01DA041174, U01DA041106, U01DA041117, U01DA041028, U01DA041134, U01DA050988, U01DA051039, U01DA041156, U01DA041025, U01DA041120, U01DA051038, U01DA041148, U01DA041093, U01DA041089, U24DA041123, U24DA041147 (ABCD Study, multiple PIs). Multiple philanthropic contributions to the HBN (Michael Milham).

## ETHICS DECLARATIONS.

The authors declare no competing financial interests.